\documentstyle[preprint,aps]{revtex}
\begin{document} 
\def\qdify{\partial_y^{(q)}}
\def\qdift{\partial_t^{(q_t)}}
\def\qdifx{\partial_x^{(q)}}
\def\be{\begin{equation}}
\def\ee{\end{equation}}
\draft 
\title{Non-commutative geometry and irreversibility} 
\author{ Ay\c se Erzan$^{1,2}$ and Ay\c se Gorbon$^1$} 
\address{$^1$Department of Physics, Faculty of  Sciences and 
Letters\\ 
Istanbul Technical University, Maslak 80626, Istanbul, Turkey 
and\\ 
$^2$T\"UBITAK Research  
Institute for Basic Sciences,\\ 
P.K. 6, \c Cengelk\"oy, Istanbul 81220, Turkey 
} 
\date{\today} 
\maketitle 
\begin{abstract} 
A kinetics built upon $q$-calculus, the calculus of discrete
dilatations, is shown to describe diffusion on a hierarchical
lattice. The only
observable on this ultrametric space is the  ``quasi-position"
whose eigenvalues
are the levels of the hierarchy, corresponding to the volume of
phase space available to the system at any given time.  Motion
along the 
lattice of quasi-positions is irreversible.
\end{abstract}
\pacs{5.20.Dd, 5.70.Ln}
\section{Introduction} 
The study of systems that are symmetric under dilatations rather
than translations have been with us for a long time.  The field
of critical phenomena has been a breeding ground for useful
scaling ideas.  Fractal geometry \cite{Mandelbrot} has provided
us with a suitable language with which to describe systems with
affine symmetries.  It has been recently demonstrated \cite{AEJP}
that, although ordinary derivatives of fractal or multifractal
distributions may be nowhere defined, the finite $q$-derivative
\cite{Jackson} provides a natural extension of the derivative  to
systems with discrete dilatation symmetries, and the $q$-integral
\cite{Jackson1} provides the requisite tool for integrating along
a discrete path in scale space \cite{Erzan}.

An appropriate language in which to describe the kinetics and
dynamics of motion on such spaces, however, has not yet been
sufficiently elaborated.  One may think that a periodic lattice
on the logarithmic scale would play here the same role as the
linear chain does with respect to motion on a discrete,
translationally invariant space.  However, it is easy to see that
progress on the logarithmic lattice is not symmetric with respect
to a simultaneous reflection and time reversal, and corresponds
to very different physics.  It is the purpose of this paper to
explore this asymmetry and to show that taking a statistical view
point and associating the kinetics of a point on the logarithmic
lattice with the motion of a representative point in phase space,
leads naturally to the arrow of time one encounters 
in statistical physics
\cite{Lebowitz}.

The paper is organized as follows.  In the next section, we will
briefly recall the work of Dimakis and M\"uller-Hoissen
\cite{Dimakis} relating the so called $q$-deformed quantum
mechanics \cite{qmech} to quantum mechanics on a discrete lattice
and then make a different, and we claim more natural, choice for
the operators, to describe a different physics.  In this
description, the energy and momentum are no longer observables,
nor are they conserved. Instead, we define a ``quasi-position''
operator, and show, in section 3, that this tells us the volume
in phase space over which the probability distribution of the
representative point of our system is spread.  In section 4 we
discuss connections with other recent work.

\section{A Hamiltonian system on a hierarchical space}
It has been demonstrated by Dimakis and M\"uller-Hoissen
\cite{Dimakis} that $q$-calculus \cite{Jackson,Jackson1}
can be obtained from discrete calculus on a lattice by an
exponential coordinate transformation. Under this transformation
discrete translations go over to discrete dilatations. 
The $q$-deformed commutation relations obeyed by the transformed
variables and their $q$-derivatives lead to $q$-deformed quantum
mechanics \cite{qmech}.  In this way, $q$-deformed quantum
mechanics has been given an interpretation in terms of quantum
mechanics on a lattice.

Let us recall the definition of the $q$-derivative
~\cite{Jackson,Dimakis,Vil-Klim,basic},
\begin{equation}
\partial_y^{(q)} f(y) \equiv 
{f(qy)-f(y) 
\over (q-1)y}\;\; ,\label{qdif1}\end{equation} 
and
\begin{equation}
\overline{\partial}_y^{(q)} f(y) \equiv 
{f(q^{-1}y)-f(y) 
\over (q^{-1}-1)y}\;\; .\label{qdif2}\end{equation} 
where the subscipt indicates the variable with respect to which
the derivative is to be taken. It is easy to see that
\cite{Dimakis} 
these
operators can be obtained from the discrete partial derivatives
\begin{eqnarray}
\tilde \partial_x^{(a)} f(x) \equiv {1 \over a} [f(x+a)-
f(x)] \label{adif1} \\
\overline{\tilde \partial}_x^{(a)} f(x) \equiv {1 \over a} [f(x)-
f(x-a)] \label{adif2}
\end{eqnarray}
under the coordinate transformation
\begin{eqnarray}
y&=q^{{x\over q-1}}\\
q&=1+a\;\;.\end{eqnarray}
With $x=\ell a$, one  has $y=q^\ell$, since $q-1=a$. 
Thus, this coordinate transformation takes the one-dimensional
lattice with lattice spacing $a$, to another lattice which has
spacing $q$ on the logarithmic scale (Fig.1a).
It is  useful to define discrete translation operators on the
lattice in $x$-space, in terms of which the difference operator 
can be
expressed; under the change of variables these go over to the
discrete dilatation operators such that $(A_y^{(q)}f)(y)=f(qy)$
and $(\overline{A}_y^{(q)} f)(y)=f(q^{-1}y),$ with, 
\begin{equation}
A_y^{(q)}\equiv 1+(q-1)Y\partial_y^{(q)}\label{dil}
\end{equation}
and 
\begin{equation}
\overline{A}_y^{(q)}\equiv 1-(q-1){1\over
q}Y\overline{\partial}_y^{(q)}\;\;.
\end{equation}

If the position operator $X$ is defined as multiplication by $x$,
we notice that it is self-adjoint, and therefore can be
identified with an observable, while the one sided 
(\ref{adif1},\ref{adif2}) difference operators $\tilde
\partial_x^{(a)}$ and  $\overline{\tilde\partial}_x^{(a)}$  are
not. On the other hand, Dimakis and M\"uller-Hoissen
\cite{Dimakis} define the momentum and Hamiltonian operators in
$x$-space via self-adjoint linear combinations of these
operators.  The momentum and Hamiltonian thus obtained satisfy
the Heisenberg equations of motion; however, the usual canonical
commutation relation is altered.  Nevertheless, interpreting this
commutator as giving the time evolution operator for a free
particle they are able to write down the ``Schr\"odinger"
equation
and find its solutions. 

Going over to the transformed space, with $Y$ straightforwardly
implying multiplication by $y$, the following ``$q$-deformed''
commutation relations,
\begin{equation}
[\partial_y^{(q)},Y]_q \equiv \partial_y^{(q)}\, Y - q\, Y 
\partial_y^{(q)}=1\label{qcom}\end{equation}
and 
\begin{equation} [\overline{\partial}_y^{(q)},Y]_{q^{-
1}}=1\label{q-com}\end{equation}
\begin{equation}
[\overline{\partial}_y^{(q)},\partial_y^{(q)}]_q=0\;\;,
\end{equation}
hold. The transformed momentum and Hamiltonian operators remain
hermitian. However, they satisfy Heisenberg's
equations of motion with the ordinary definition of the
commutator and not with the deformed definition.
To be able to give an
intepretation of the physics, one has to transform back to the
linear lattice. 

We would now like to propose a different choice for the momentum 
operator.
Notice that there is a kind of democracy between the  right
and left difference operators (\ref{adif1},\ref{adif2}), which
makes it
natural for the (self-adjoint) momentum operator on the discrete
lattice to be defined~\cite{Dimakis} as, 
$(\tilde\partial_x^{(a)}
+\overline{\tilde\partial}_x^{(a)})/(2i)$
but this democracy does not hold between $\partial_y^{(q)}$ and
$\overline{\partial}_y^{(q)}$ which describe processes at {\it
different
scales}. On the linear chain, exactly one unit is added 
to an interval everytime a step is made to the right
wherever one may be on the chain.  
However, when $\ell$ is increased by unity in $y$
space, the size of the interval which is certain to include the 
origin increases by 
$(q-1)q^{\ell}$.
We will therefore deliberately allow the
momentum not to be an observable. This gives us the
freedom to associate the momentum operator directly with the 
$q$-derivative (\ref{qdif1})
\begin{equation}
P_q=-\,i\;\partial_y^{(q)}\;\;.\label{P}
\end{equation}
Now we consider the ordinary commutator of $Y$ and $P_q$ rather
than the $q$-deformed one as in (\ref{qcom}).  We find that the
canonical commutation relation becomes,
\begin{equation}
[P_q,Y]=\,-\,i\;A_y^{(q)}\;\;.\label{commut}\end{equation}
Comparing this with $[P_q,Y]=-iT$, we find that the time
evolution operator, $T$, which is defined by 
\begin{equation}
Tf(y,t)\equiv f(y,q_t \, t)\label{T}
\end{equation}
is identical with the dilatation operator,
\begin{equation}
T=A_y^{(q)}\;\;. \label{TA}
\end{equation}
Clearly, $q_t$ need not be equal to $q$; in fact one may define
the ``dynamical exponent'' via 
\begin{equation}
q_t=q^\zeta\;\;.\label{zeta}\end{equation}
We may now write down the deformed ``Schr\"odinger equation'' and
thereby identify the Hamiltonian operator from,
\begin{equation}
i \,\partial_t^{(q_t)} f(y,t)=H_q f(y,t)\;\;.
\label{Schr}\end{equation}
Using the definitions (\ref{qdif1}, \ref{T}) one readily has
\begin{equation}
H_q= i\, {T-1 \over (q_t -1) t}\;\;,\label{Hamilt1}\end{equation}
or, with (\ref{TA}) and (\ref{dil}), and defining the imaginary
time $t=i \tau$, 
\begin{equation}
H_q={ (q-1) Y\partial_y^{(q)} \over (q_t -1) \tau}\;\;.
\label{Ham}\end{equation}
This operator is also non-hermitian, so that the energy is not an
observable, neither is it a constant of the motion; the
Hamiltonian depends explicitly on time.  Since $[H,P_q]\neq 0$, 
the momentum is not conserved either. 
Thus we see that with this choice for the momentum operator, the
{\it conventional} commutation relations together with the same
coordinate transformation leads once more to a non-conventional
mechanics.  We will show below that here, the motion of a ``free
particle'' corresponds to diffusion on a hierarchical lattice.

The constant prefactor in Eq.(\ref{Ham}) 
may be written
as the inverse of a ``basic number''~\cite{basic},
\begin{equation}
[\zeta]_q\equiv {q^\zeta -1 \over q-1}\;\;,\label{basic}
\end{equation}
where $\zeta$ is the dynamical exponent defined in (\ref{zeta}).
This dynamical exponent $\zeta$, which tells us how time scales
with the
distance, takes the value of 2 in the case of diffusion on
Eucliean space. With these definitions, the Hamiltonian operator
becomes  
\begin{equation}
H_q=\,-\, {1 \over [\zeta]_q} {Y\;P_q\over
t}\;\;,\label{Ham1}\end{equation}
which has the right ``dimensions'' for being an energy.

The solutions of the ``Schr\"odinger equation'' (\ref{Schr}) can
be found by making a seperation of variables, so that
$f(y,t)=g(y) h(t)$.  Then, using (\ref{Ham1}) and (\ref{P}), one
has
\be
{t\qdift h(t) \over h(t)} = {1\over [\zeta]_{q_t}} {Y\qdify g(y)
\over g(y)}\;\;.\ee
Setting both sides of the equation equal to a constant, $C$,
gives,
\be
{t\qdift h(t) \over h(t)} =C\;\;, \label{sepvar1}\ee
and
\be
 {Y\qdify g(y) \over g(y)}=[\zeta]_{q_t} C\;\;.\label{sepvar2}\ee
The solutions to these equations are given in terms of
homogeneous functions, namely power laws, up to multiplication by
oscillatory functions, 
\be
h(t)=F_{q_t}(t)t^\psi\;\;,\label{sepvar3}\ee
\be g(y)=F_q(y) y^\chi\;\;.\label{sepvar4}\ee
From (\ref{zeta}), we find
$[\zeta]_q [\psi]_{q_t}=[\zeta\, \psi]_q$. 
On the other hand, from (\ref{sepvar2}) and (\ref{sepvar4}), we
have $[\chi]_q=[\zeta]_q [\psi]_{q_t}$, whence, $\chi = \zeta
\psi$. For finiteness as $t\to \infty$, $\chi, \;\psi<0$.

The oscillatory amplitudes multiplying the power laws in
(\ref{sepvar3},\ref{sepvar4}) must satisfy $F_r(ru)=F_r(u)$ so that

$\partial_u^{(r)} F_r(u)=0$.  Such functions periodic in the
logarithm of their arguments can be 
expressed in terms of 
the Jackson integral \cite{Jackson1,Vil-Klim} from $0$ to
$\infty$
\begin{eqnarray}
F_r(u)&=&\int_{0 u}^{\infty u} \phi(v)D_v^{(r)}\\
 &\equiv& \int_0^{u} 
{\phi(v)\over v^{1+\omega} }D_v^{(r)}
+\int_{u}^\infty {\phi(v) \over v^{1+\omega}} D_v^{(r)}\\
&=&(1-r)u^{-\omega} \sum_{k=-\infty}^{\infty}r^{-k\omega} 
\phi(r^ku)\;\;,
\end{eqnarray}
where we have used \cite{AEJP} the notation $D_v^{(q)}$ for 
the $q$-differential of $v$; $\phi(v)$ is an arbitrary periodic
function,  which vanishes  at the origin together with its first
$n_0$ derivatives, 
$n_0$ being the smallest integer $> \omega$, and $\omega >0$.
Notice that since $k$ ranges over all 
positive {\it and} negative values, $r$ here may be bigger than
unity, as we have assumed $q$ to be.

Finally, the solutions of Eq.(\ref{Schr}) can be written,
\be
f_\psi(y,t) =F_q(y)F_{q_t}(t) (y^\zeta t)^\psi.\ee
These solutions are degenerate with respect to the functions
$\phi$ and the indices $\omega$ appearing in the oscillatory
amplitudes. Notice, however, that when $y$ and $t$ are only 
allowed to take discrete values such as $q^m$, $q_t^n$, with
integer  
$m$ and $n$, clearly the functions $F_q(y)$ 
and $F_{q_t}(t)$ can only take on constant values.

\section{The ``quasi-position'' operator and spreading of the 
probability distribution}

With the hermitian operator $Y$ we will associate a position-like
observable which we will call the ``quasi-position,"
\cite{Majid},
the ``quantum numbers'' $\ell$ corresponding to the highest level
so far attained by the phase point on the $y$-lattice.   
 To the motion of the phase point along the chain of 
 quasi-positions $q^\ell$, (see Fig.1a), there corresponds 
 an underlying picture 
as shown in Fig.(1b), whereby each successive quasi-position
indexed by the quantum number $\ell$ corresponds to a geometrical
increase in the number of microstates made available to the phase
point on the 
hierarchical lattice \cite{UMSdif,Schreck}.
Transitions between microstates within the same interval of size
$q^\ell$ do not change the quantum number $\ell$, i.e., the
quasi-position. To proceed from the $\ell$'th level of the
hierarchy to the next, we assume the particle has to surmount an
energy barrier of hight $R^\ell$.

A hierarchical lattice  with branching ratio $\mu$ is shown in 
Fig.(1).
The origin has been arbitrarily chosen at $y_0$. 
The regions of extent $q^{\ell}$ (or phase-space volume
$\mu^\ell$) over which the particle is successively delocalized
form a nested hierarchy, i.e., the microstates
already available at the $\ell-1$st level are subsumed by the
microstates that become available at the $\ell$th level, with an
increase in the 0-level states of 
$\mu^{\ell}(1-\mu^{-1})$.  Since going from one level to
the next involves an increase in the available phase space
volume, it implies an increase in the entropy of the system, and
therefore we expect this motion to be irreversible, which it
indeed is. 

The unnormalized state functions corresponding to  
the pure states  $\vert \ell \rangle$ of  the quasi-position
operator
are given,  again up to
multiplication by functions doubly periodic in $\ln y$ and $\ln
t$ 
with periods $\ln q$ and $\ln q_t$, by 
\begin{equation}
\epsilon_\ell(y,t)=\exp\Bigl\{-{1\over
2}\Bigl[{(y/y_0)^\zeta\over  \tau_\ell} t\Bigr]^\lambda\Bigr\}
\label{eigen}\end{equation}
where $\tau_\ell =R^\ell$ are the characteristic decay times, and
$\lambda>0$ is arbitrary.  For simplicity, we shall choose
$\lambda=1$,
but this does not at all affect the subsequent discussion.  By
(\ref{TA}), $T\epsilon_\ell(y,t)=A_y^{(q)}\epsilon_\ell(y,t)$. 
Thus, one must
have
\begin{equation}
\epsilon_\ell(y,q_t t)=\epsilon_\ell(qy, t)\;\;.\end{equation} 
Substituting from (\ref{eigen}), one finds that
 if $R \equiv q_t$, then we also have, 
\begin{eqnarray}
A_y^{(q)} \epsilon_\ell(y,t)&=\epsilon_{\ell-1}(y,t)\\
\overline{A}_y^{(q)} \epsilon_\ell(y,t)&=\epsilon_{\ell+1}(y,t)
\label{A:f}
\end{eqnarray}

The expectation value of the quasi-position operator is to be
computed using the definition of the scalar
product~\cite{Dimakis},
\begin{eqnarray}
\langle a,b \rangle_{y_0}&\equiv & \int_{0y_0}^{\infty y_0} 
{a(y)\,b(y)\over y}D_y^{(q)}\\
&\equiv& (1-q)\sum_{k=-\infty}^{\infty} a(q^ky_0)\;
b(q^ky_0)\;\;\end{eqnarray}
Here $y_0$ serves as the origin of this hierarchical lattice, and
could be chosen equal to unity.  

Defining $\langle \epsilon_\ell(y,t), y \,
\epsilon_\ell(y,t)\rangle_{y_0}
\equiv Q_\ell(t)$, we have,
\begin{equation}
 Q_\ell(t)=(1-q)y_0\sum_{k=-\infty}^\infty q^k e^{-\,q_t^{k-
\ell}t}\;\;.\label{QL}
\end{equation}
The  state functions have been chosen in such a way that they
decay
sufficiently fast for the infinite sum to converge at both ends.
  Notice that $Q_{\ell+1}(t) =Q_\ell(q_t^{-1} t)$ or $Q_\ell(t)
  =T\;Q_{\ell+1}(t)$. By (\ref{QL}) we have,
\begin{equation}
Q_{\ell+1}=(1-q)y_0\sum_{k=-\infty}^\infty q^k e^{-\,q_t^{k-1-
\ell}t}\;\;,\end{equation}
Upon redefining the dummy index to be $k^\prime=k-1$, this
gives,
\begin{equation}
Q_{\ell+1}(t)=q\,Q_\ell(t)\;\;.\end{equation}
Thus, clearly, $Q_\ell(t)=q^\ell Q_0(t)$ and the
$\epsilon_\ell(y,t)$
span a representation of the algebra generated by the
$\partial_y^{(q)},\;\overline{\partial}_y^{(q)}$ and $Y$.

Now we would like to show that the kinetics imply that a 
probability distribution initially localized within an interval 
$q^\ell$ of the origin will spread in time
in such a way that the uncertainty in the position becomes
precisely as large as the whole phase space available at time
$t$.    This means that the probability distribution is
essentially uniform over the available phase space at any given
time.

The absolute value of the uncertainty in the simultaneous
determination of the ``momentum'' and ``position'' operators can
be found as usual from the canonical commutation relation.  In
our case, from (\ref{commut},\ref{Hamilt1}) we have
\begin{eqnarray}
\vert \langle \Delta Y \Delta P_q\rangle \vert &\ge& 
\vert \langle[Y, P_q]\rangle\vert \\
&=& \vert \langle iA_y^{(q)} \rangle\vert \label{Delta:H} \\
&=&\vert  \langle-(q-1)YP_q +i\rangle \vert\;\;. \label{delta}
\end{eqnarray}
This tells us that the product of the uncertainty in the value
of the position and the momentum operators is larger than the
expectation value of their product in absolute value.  
Heuristically, one may say that, if
$Y\sim v t^{1/\zeta}$ where $v$ is some effective diffusivity, 
then the uncertainty
$\vert <\Delta Y \Delta P_q> \vert > \vert v t^{1/\zeta} p_q\vert
$, 
where $p_q$ 
is the average momentum for this $Y$ eigenstate.
Thus, 
the uncertainty in the position is as large, and increases with
time 
in the same way, as the interval over
which the particle or the phase point has travelled within the
time $t$, i.e., it is equally likely to be found anywhere within
the phase space volume it is energetically allowed to explore.

More precisely, the expectation value of $[Y,P_q]$, taken with
respect 
to the solutions of the Schr\"odinger equation, 
normalized by their scalar 
product, yields,
\begin{eqnarray}
\langle [Y,P_q]\rangle &=& i\langle [(q-1)Y\qdify +1]\rangle\\
                       &=& i\Bigl[(q-1) {q^{\zeta \psi} -1
\over q-1} +1\Bigr]=iq^\chi\;\;.
                       \end{eqnarray}
With $q^\zeta=q_t$, this yields,
\begin{equation}
\vert <\Delta Y \Delta P_q> \vert \ge 
q^\chi=q_t^{\psi}\;\;.\label{qtn}
\end{equation}

\section{Discussion and connection with $q$-statistics}
We note that $[Y ,P_q] = iT$ is an operator itself, rather than
a constant, and in fact is proportional to the time-evolution
operator.  By (\ref{TA}) and (\ref{A:f}), 
taking the expectation value of
this expression between the states $\vert \ell\rangle$
gives us $\langle \epsilon_\ell, \epsilon_{\ell-1}\rangle$,
which may be
interpreted as a transition probability between the states
$\vert \ell -1 \rangle$
and $\vert \ell\rangle$.  Again this is telling us that the
uncertainty
increases as a function of the leakage of the phase point to
larger and larger regions of the phase space. It is  interesting
to remark that our Schr\"odinger equation (\ref{Schr}, \ref{Ham})
involves, on the RHS, only the first derivative with respect to
position, in accordance with the fact that diffusion on the
hierarchical lattice corresponds to simply a drift with respect
to the quasi-position.  This makes the Schr\"odinger equation
resemble the Fokker-Planck equation rather than the diffusion
equation.

In statistical physics, hierarchical lattices have arisen
recently
in the anomalous relaxation of spin glasses \cite{Rammal,Cyrano},
transport in  
random media \cite{Alexander}
and fully developed turbulent media
\cite{UMSdif} as realizations of ultrametric spaces \cite{UMS}. 
They consist of a hierarchy
of nested intervals (see Fig.1), and one may associate a
geometrical progression of spatial (and/or temporal) scales with
the different levels of the hierarchy.  Diffusion on ultrametric
spaces have been thoroughly studied (see 
\cite{UMSdif,Schreck,Rammal,Cyrano,Alexander,Sibani,UMS} 
and references therein) by other methods, including the
renormalization group.  Here we have pointed out that on a
lattice with equal spacing on the logarithmic scale, a natural
choice for
the position (``quasi-position'') and  momentum operators,
together with the canonical commutations relation yields a
kinetics that can be understood in terms of diffusion on an
underlying ultrametric space.  The motion to which this 
non-conventional kinetics corresponds is irreversible, with
an explicit violation of time reversal symmetry resulting from
the spreading with time of the probability distribution over a
larger and larger volume of the phase space~\cite{Lebowitz}.

We would like to mention that Dimakis and Tzanakis\cite{Dimakis2}
have also recently given an alternative 
 description of the kinetics of open systems, 
built upon the assumption that observables are now defined 
on a manifold with non-commutative geometry.  In this way, they 
recover  the 
non-conventional calculus obeyed by stochastic differentiation 
(It\^o's calculus),  without making 
any uncontrolled approximations with respect to the microscopic 
Hamiltonian dynamics of the system.  The relationship between our
approaches, however, will be the subject of a different study.

Finally, we would like to make a connection with recent work on
random sets and $q$-distributions. It has been remarked
by Ar\i k et al.~\cite{Arik} that the basic number $[n]_q$ with 
$q=1-1/M<1$ is 
 the average
number of distinct elements in a set which is contructed in $n$
steps by making random draws from a source set with infinitely
many elements of which there are $M$ distinct kinds.  
In our case, $q>1$,  which is complementary to that considered by
Ar\i k 
et al. 
The spreading of the distribution in the phase space of our
system 
extends at each step by $(1-1/q)q^m=(q-1)q^{m-1}$, so that the
total  volume explored in $n$ steps is precisely $1+ (q-1)
\sum_1^{n}q^{m-1}=q^n$.  

{\bf Acknowledgements}

It is a pleasure to thank Metin Ar\i k for several useful
discussions and 
for making their results available to us prior to publication. 
One of us (AE) would like to acknowledge partial support by the
Turkish Academy of Sciences.  AG is partially supported under the
BDP program by the Technical and Scientific Research Council of
Turkey (TUBITAK).

\vfill
\eject

\vfill
\eject

{\bf Figure Caption}

The hierarchical space on which the quasi-position operator $Y$
is defined.  The ``quantum number'' $\ell$ corresponds to the
highest node which the particle has so far surmounted, 
with an associated energy barrier of height $R^\ell$, 
thereby
being delocalized over a region of size $q^{\ell}y_0$.  
We will define the distance $d$ between two states such that 
$q^d y_0$ is the smallest interval which contains both of them.
Thus,
the quasi-position 
$\ell$ is the upper bound on the  distance $d$ to the origin at
any
given time. Note that we allow negative values of $d$; this is 
useful since we have taken each state at level 0 to be itself 
infinitely divisible, so that the tree is scale invariant under 
all dilatations. Transitions between  microstates whose
distances to the origin are $d\leq \ell$ do not affect
the ``quasi-position'' $\ell$. The  tree shown in the figure has
branching ratio $\mu=2$.
\vfill
\eject
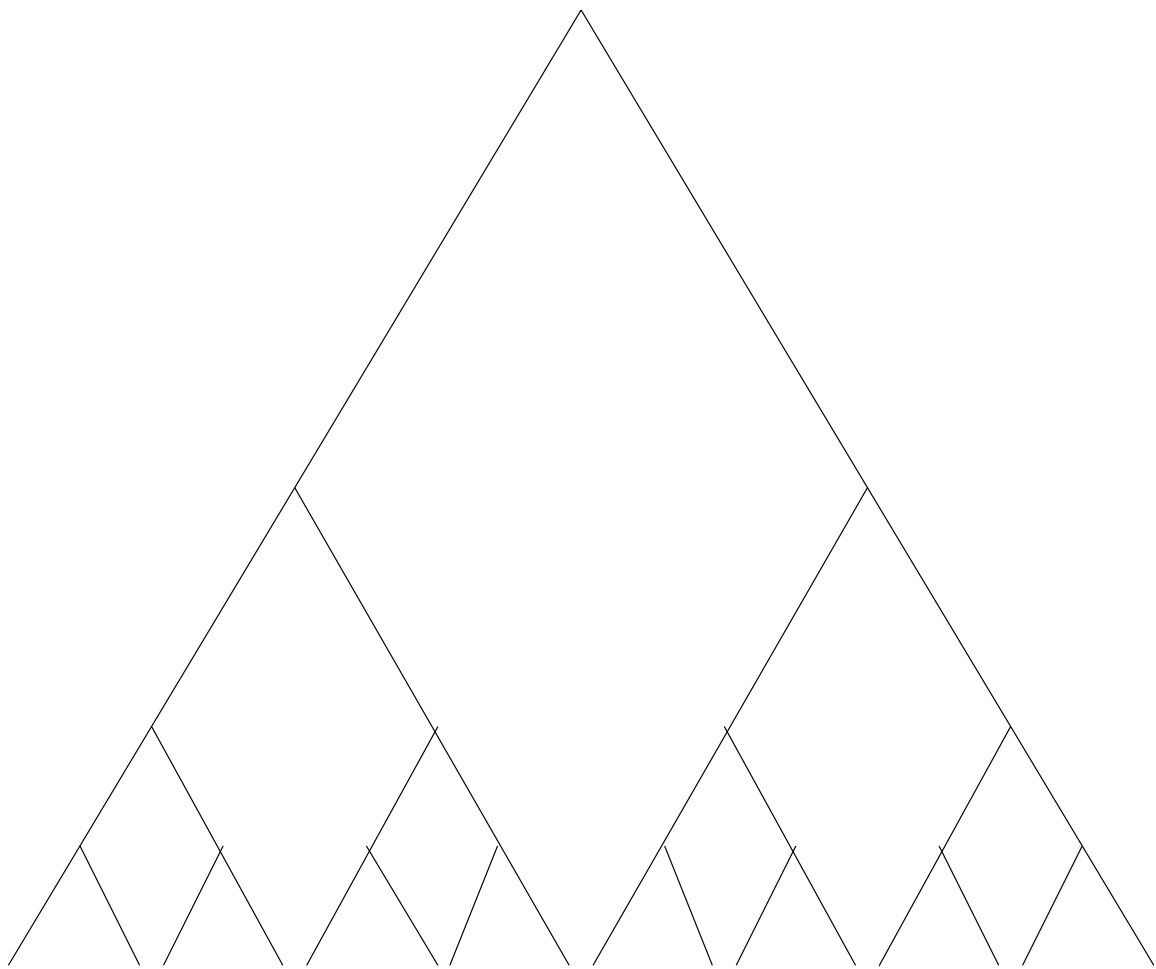
\end{document}